\documentclass[12pt]{article}
\def\be{\begin{equation}}
\def\ee{\end{equation}}
\def\bea{\begin{eqnarray}}
\def\eea{\end{eqnarray}}
\def\ba{\begin{array}}
\def\ea{\end{array}}

\begin{document}
 
\title{Singlet GB contributions in the chiral constituent quark model}
\author{Harleen Dahiya$^a$, Manmohan Gupta$^a$ and J.M.S. Rana$^{b,c}$     \\
{\small {\it $^a$Department of Physics, Centre of Advanced Study in Physics,}} \\
{\small{\it Panjab University, Chandigarh-160 014, India.}} \\
{\small{\it $^b$ Abdus Salam International Centre for Theoretical Physics, Trieste Italy.}}\\
{\small{\it $^c$ Department of Physics, H.N.B. Garhwal University, SRT Campus }}\\
 {\small{\it Badshahithaul(Tehri-Garhwal), India.}}}
\maketitle

\begin{abstract}

The implications of the latest data pertaining to 
$\bar u-\bar d$ asymmetry and the spin polarization functions on the 
contributions of singlet Goldstone Boson
$\eta'$ within $\chi$CQM with configuration mixing for explaining the
``proton spin problem''
have been investigated. It is found that the present data favors
smaller values of the coupling of  singlet Goldstone Boson $\eta'$ as
compared to the corresponding contributions from $\pi$, $K$ and $\eta$
Goldstone bosons. It seems that a small non-zero value of the coupling
of $\eta'$ ($\zeta \neq 0$) is preferred over $\zeta=0$
phenomenologically. 

\end{abstract}

The chiral constituent quark model ($\chi$CQM), as formulated by
Manohar and Georgi \cite{{manohar},{wein}}, has recently got good deal of
attention \cite{{eichten},{cheng},{song},{johan}}
as it is successful in not only explaining the ``proton spin crisis''
\cite{emc,smc,abe,hermes}
through the emission of a Goldstone boson (GB) but is also able to
account for the $\bar u-\bar d$ asymmetry
{\cite{{nmc},{e866},{GSR}}}, existence of significant strange 
quark content $\bar s$ in the nucleon,  various
quark flavor contributions to the proton spin {\cite{eichten}}, baryon
magnetic moments {\cite{{eichten},{cheng}}}
and hyperon $\beta-$decay parameters etc..

Recently, it has been shown that configuration mixing generated by
spin-spin forces \cite{{dgg},{Isgur},{yaouanc}}, 
known to be compatible with the $\chi$CQM
\cite{{riska},{chengspin},{prl}}, improves the predictions of 
$\chi$CQM regarding the quark distribution functions and the
spin polarization functions \cite{hd}. Further, $\chi$CQM with
configuration mixing (henceforth to be referred as $\chi$CQM$_{{\rm
config}}$ ) when coupled with the quark sea polarization and orbital angular
 omentum (Cheng-Li mechanism \cite{{cheng1}}) as well as
``confinement effects'' is able to give an excellent fit 
\cite{hd} to the octet magnetic moments and 
a perfect fit for the violation of Coleman Glashow sum rule \cite{cg}.

Cheng and Li \cite{cheng} realized that the
key to understand the ``proton spin problem'' \cite{psp}, 
within the Manohar
and Georgi formalism of $\chi$CQM \cite{manohar}, lies in
generating an appropriate quark sea in the proton through the
{\it chiral symmetry breaking} mechanism. 
The basic process in the $\chi$CQM is the
emission of a GB by a constituent quark which further splits into a $q
\bar q$ pair, for example,  
\be
  q_{\pm} \rightarrow {\rm GB}^{0}
  + q^{'}_{\mp} \rightarrow  (q \bar q^{'})
  +q_{\mp}^{'}\,, \label{basic}
\ee
where $q \bar q^{'}  +q^{'}$ constitute the ``quark sea''
  \cite{{cheng},{song},{johan}}. 
The effective Lagrangian describing interaction between quarks and a nonet of
GBs, consisting of octet and a singlet, can be expressed as
\be
{\cal L}= g_8 {\bf \bar q}\Phi {\bf q} +
g_1{\bf \bar q}\frac{\eta'}{\sqrt 3}{\bf q}=
g_8 {\bf \bar q}\left(\Phi+\zeta\frac{\eta'}{\sqrt 3}I \right) {\bf q}\,,
\ee
where $\zeta=g_1/g_8$, $g_1$ and $g_8$ are the coupling constants for the 
singlet and octet GBs, respectively, $I$ is the $3\times 3$ identity
matrix and 
\be 
q =\left( \ba{c} u \\ d \\ s \ea \right)\,. 
\ee 
The GB field which includes the octet and the singlet GBs is written as
\bea 
 \Phi = \left( \ba{ccc} \frac{\pi^0}{\sqrt 2}
+\beta\frac{\eta}{\sqrt 6}+\zeta\frac{\eta^{'}}{\sqrt 3} & \pi^+
  & \alpha K^+   \\
\pi^- & -\frac{\pi^0}{\sqrt 2} +\beta \frac{\eta}{\sqrt 6}
+\zeta\frac{\eta^{'}}{\sqrt 3}  &  \alpha K^0  \\
 \alpha K^-  &  \alpha \bar{K}^0  &  -\beta \frac{2\eta}{\sqrt 6}
 +\zeta\frac{\eta^{'}}{\sqrt 3} \ea \right)\,. \eea

SU(3) symmetry breaking is introduced by considering
$M_s > M_{u,d}$ as well as by considering
the masses of GBs to be nondegenerate
 $(M_{K,\eta} > M_{\pi})$ {\cite{{song},{johan},{cheng1}}}, whereas 
  the axial U(1) breaking is introduced by $M_{\eta^{'}} > M_{K,\eta}$
{\cite{{cheng},{song},{johan},{cheng1}}}.
The parameter $a(=|g_8|^2$) denotes the transition probability
of chiral fluctuation
of the splittings  $u(d) \rightarrow d(u) + \pi^{+(-)}$, whereas 
$\alpha^2 a$, $\beta^2 a$ and $\zeta^2 a$ respectively 
denote the probabilities of transitions of
$u(d) \rightarrow s  + K^{-(0)}$, $u(d,s) \rightarrow u(d,s) + \eta$,
 and $u(d,s) \rightarrow u(d,s) + \eta^{'}$.

Recently, it has been pointed out that the new measurement of both the
$\bar u/\bar d$ asymmetry as well as  $\bar u-\bar d$ asymmetry by the NuSea
Collaboration \cite{e866} may not require substantial contribution of
$\eta'$ \cite{johan}. 
As the contribution of $\eta'$ not only has important
implications for the $\chi$CQM but also has a deeper significance
for axial U(1) anomaly as well as nonperturbative aspects of QCD including
the effects of gluon anomaly on the spin polarizations \cite{bass}, 
it therefore becomes interesting to understand the extent to which its
contribution is needed in the $\chi$CQM to fit the data
pertaining to the ``proton spin problem''.

The purpose of the present communication is to phenomenologically
estimate the contribution of $\eta'$ GB by carrying out a fine grained
analysis  of  ``proton spin problem''
within $\chi$CQM$_{{\rm config}}$ which also includes the implications of
the latest E866 data. Further, it would be interesting to fine tune
the contribution of $\eta'$, expressed through the parameter $\zeta$, by
studying its implications on spin polarization functions and quark 
distribution functions.

The details of $\chi$CQM$_{{\rm config}}$ have already been discussed in 
Ref. \cite{hd}, however to facilitate the discussion as 
well as for the sake of readability of the manuscript, some essential details of
$\chi$CQM with configuration mixing have been presented in the
sequel. 
As has already been discussed that 
spin-spin forces generate
configuration mixing \cite{{dgg},{Isgur},{yaouanc}}
which effectively leads to modification
of the spin polarization functions \cite{hd}.
The most general configuration mixing 
in the case of octet baryons \cite{{Isgur},{yaouanc},{full}} 
can be expressed as
\bea
|B \rangle&=&\left(|56,0^+\rangle_{N=0} \cos \theta +|56,0^+ \rangle_{N=2}  
\sin \theta \right) \cos \phi \nonumber \\
&&+  \left(|70,0^+\rangle_{N=2} \cos \theta^{'} +|70,2^+\rangle_{N=2}  
\sin \theta^{'} \right) \sin \phi\,, \label{full mixing}
\eea
where $\phi$ represents the $|56\rangle-|70\rangle$ mixing,
$\theta$ and  $\theta^{'}$ respectively correspond to the mixing 
among $|56,0^+\rangle_{N=0}-|56,0^+ \rangle_{N=2}$ states and
$|70,0^+\rangle_{N=2}-|70,2^+\rangle_{N=2}$ states.
For the present purpose, it is adequate
{\cite{{yaouanc},hd,{mgupta1}}} to consider the mixing only between
$|56,0^+ \rangle_{N=0}$ and the $|70,0^+\rangle_{N=2}$ states and 
the corresponding ``mixed'' octet of baryons is expressed as
\begin{equation}
|B\rangle 
\equiv \left|8,{\frac{1}{2}}^+ \right> 
= \cos \phi |56,0^+\rangle_{N=0}
+ \sin \phi|70,0^+\rangle_{N=2}\,,  \label{mixed}
\end{equation} 
for details of the  spin, isospin and spatial parts of the 
wavefunction,  we  refer the reader to reference {\cite{{yaoubook}}. 

To study the variation of the $\chi$CQM parameters and the role of
$\zeta$ in obtaining the fit, one needs to formulate the
experimentally measurable quantities having implications for these
parameters as well as dependent on the unpolarized
quark distribution functions and the  spin polarization functions.
We first
calculate the spin polarizations and the related quantities which
are affected by the ``mixed'' nucleon.
The spin structure of a nucleon is defined as
\cite{{cheng},{song},{johan}} 
\be
\hat B \equiv \langle B|N|B\rangle,
\ee
where $|B\rangle$ is the nucleon wavefunction defined in Eq. (\ref{mixed})
and $N$ is the number operator given by
\be
 N=n_{u^{+}}u^{+} + n_{u^{-}}u^{-} +
n_{d^{+}}d^{+} + n_{d^{-}}d^{-} +
n_{s^{+}}s^{+} + n_{s^{-}}s^{-}\,, 
\ee
where $n_{q^{\pm}}$ are the number of 
$q^{\pm}$ quarks. 
The spin structure of the ``mixed'' nucleon, defined through the
 Eq. (\ref{mixed}), is given by
\be
 \left\langle 8,{\frac{1}{2}}^+|N|8,{\frac{1}{2}}^+\right\rangle=\cos^2 \phi
\langle 56,0^+|N|56,0^+\rangle+\sin^2 \phi\langle70,0^+|N|70,0^+\rangle. \label{spinst}
\ee
The contribution to the proton spin in $\chi$CQM$_{{\rm config}}$, given by the
spin polarizations defined as $\Delta q=q^+-q^-$, can be written as 

\bea
   \Delta u &=&\cos^2 \phi \left[\frac{4}{3}-\frac{a}{3}
   (7+4 \alpha^2+ \frac{4}{3} \beta^2
   + \frac{8}{3} \zeta^2)\right] \nonumber \\
   &&+ \sin^2 \phi \left[\frac{2}{3}-\frac{a}{3} (5+2 \alpha^2+
  \frac{2}{3} \beta^2 + \frac{4}{3} \zeta^2)\right],\\
  \Delta d &=&\cos^2 \phi \left[-\frac{1}{3}-\frac{a}{3} (2-\alpha^2-
  \frac{1}{3}\beta^2- \frac{2}{3} \zeta^2)\right]  \nonumber \\
&&+ \sin^2 \phi \left[\frac{1}{3}-\frac{a}{3} (4+\alpha^2+
  \frac{1}{3} \beta^2 + \frac{2}{3} \zeta^2)\right], \\
   \Delta s &=& -a \alpha^2\,.
\eea

After having formulated the spin polarizations of various  quarks, 
we consider several measured quantities 
which are expressed in terms of the above mentioned spin  polarization 
functions.
The quantities usually calculated in the $\chi$CQM are
the flavor non-singlet components $\Delta_3$
and $\Delta_8$, obtained from the neutron $\beta-$decay and the weak
decays of hyperons respectively. These can be related to  Bjorken sum
rule \cite{bjorken} and the 
Ellis-Jaffe sum rule \cite{ellis} as
\bea
BSR: &&\Delta_3= \Delta u-\Delta d\,, \label{BSR}\\
EJSR:&&\Delta_8= \Delta u+\Delta d-2 \Delta s\,. \label{EJSR}
\eea

Another quantity which is usually evaluated is the flavor singlet 
component of the total quark spin
content defined as 
\be 
2 \Delta \Sigma= \Delta_0= \Delta u+\Delta  d+\Delta s\,.
\ee

Apart from the above mentioned spin polarization 
we have also considered the quark distribution functions
which have implications for $\zeta$ as well as for other $\chi$CQM
parameters. For example, the
antiquark flavor contents of the ``quark sea''
can be expressed as  \cite{{cheng},{song},{johan}} 
\be
\bar u =\frac{1}{12}[(2 \zeta+\beta+1)^2 +20] a\,,~~~
\bar d =\frac{1}{12}[(2 \zeta+ \beta -1)^2 +32] a\,, ~~~
\bar s =\frac{1}{3}[(\zeta -\beta)^2 +9 {\alpha}^{2}] a\,,
\ee
and
\be
u-\bar u=2\,, ~~~d-\bar d=1\,, ~~~ s-\bar s=0\,.
\ee

The deviation of
Gottfried  sum rule \cite{GSR} is expressed as 
\be
I_G =\frac{1}{3}+\frac{2}{3} \int_0^1 {[\bar u(x)-\bar
d(x)] dx}=0.254\pm 0.005\,. \label{devgsr}
\ee
In terms of the symmetry breaking parameters $a$, $\beta$ and
$\zeta$, this deviation is given as
\be
\left[I_G-\frac{1}{3}\right]=
\frac{2}{3}\left[\frac{a}{3}( 2 \zeta+ \beta-3)\right]\,. \label{zeta}
\ee
Similarly, $\bar u/\bar d$ {\cite{e866,baldit}} measured 
through  the ratio of  muon pair production
cross sections  $\sigma_{pp}$ and $\sigma_{pn}$, is expressed in the 
present case as follows 
\be
\bar u/\bar d=\frac{(2 \zeta +\beta +1)^2+20}{(2 \zeta+ \beta-1)^2 +32}\,.
\ee
Some of the important quantities depending on the quark distribution functions which
are usually discussed in the literature are as follows
\be
f_q=\frac{q+\bar q}{[\sum_{q} (q+\bar q)]}\,, ~~~~f_3= f_u-f_d\,, 
~~~~f_8= f_u+f_d-2 f_s \,.
\ee

The $\chi$CQM$_{{\rm config}}$ involves five parameters: $a$, $\alpha$,
$\beta$, $\zeta$ and $\phi$. 
Before carrying out the detailed analysis involving quantities which
are dependent on $\zeta$, to begin with we have fixed some of the $\chi$CQM
parameters. The mixing angle $\phi$ is fixed from the
consideration of neutron charge radius
\cite{{yaouanc},{full},{neu charge}}. 
It has been shown \cite{{cheng},{johan}} that to fix the
violation of Gottfried sum rule \cite{GSR}, we have to  consider the
relation
\be 
\bar u-\bar d=\frac{a}{3}(2 \zeta+\beta-3)\,,
\ee
which constraints the parameters $a$, $\zeta$ and $\beta$ when the
data pertaining to $\bar u-\bar d$ asymmetry \cite{e866} is used.
The parameters $\alpha$ and $\beta$ suppress the emission of $K$
and $\eta$ as compared to that of pions as these strange quark carrying
GBs are more massive than the pions. However, because of the very
small mass difference between them, the  suppression factors
$\alpha$ and $\beta$ are taken to be equal.
In Table \ref{input}, we summarize the input 
parameters and their values.

In Table \ref{spin}, we have presented the various spin dependent
phenomenological quantities which are affected by the
variation of the symmetry breaking parameters. 
In the table, to highlight the particular values of $a$ and $\zeta$,
we have presented the results for their different values.
A general look at the table shows that the results of all 
the quantities affected by the inclusion of $\zeta$ get 
improved in the right direction for lower values of $\zeta$.
In fact, for the case of $a=0.13$ and $\zeta=-0.10$, we are able to
get a perfect fit for $\Delta_3$ and  $\Delta_8$.

Further, the results corresponding to quark distribution
functions having implications for the symmetry breaking parameters
have been presented in Table \ref {quark}.
In general both for $\zeta=0$ and $\zeta=-0.10$, we are able to obtain an
excellent fit, however in the case of $\bar u-\bar d$,  $\bar u/\bar
d$ and $f_3/f_8$, the non-zero (small) value of $\zeta$ gives a better
fit than $\zeta=0$.

A closer scrutiny of  the table reveals several interesting points.
$\Delta_3$ and $\Delta_8$ from Table \ref{spin} as well as $f_3/f_8$
from Table \ref{quark} perhaps suggest that a small non-zero value of
$\zeta$ gives a better fit than the zero value of $\zeta$. In the case
of $\Delta \Sigma$ (Table \ref{spin}), it seems that $\zeta=0$ is a
preferred value. However, as has been discussed earlier in $\chi$CQM
\cite{hdgluon} that the flavor singlet component of
the spin of proton $\Delta \Sigma$ receives contributions from
various sources such 
as gluon polarization and gluon angular momentum, therefore, we cannot
conclude that $\zeta=0$ is preferred over $\zeta \neq 0$.  
In this context, we would like to
mention that the above contribution of $\eta'$ is in agreement with
the experimental value of $\Delta \Sigma$ in case we consider the
contribution of the effects of gluon polarization and gluon angular
momentum through gluon anomaly \cite{hdgluon}.

The results corresponding to small values of $\zeta$ including
$\zeta=0$ clearly show better overlap with the data after the latest
$\bar u-\bar d$ asymmetry measurement \cite{e866}. In the $\chi$CQM, it is
difficult to think of a mechanism wherein the contribution of $\eta'$
or the ninth GB becomes zero. However, a small value of $\zeta$ looks
to be in order from phenomenological considerations pertaining to the
different GBs. For example, in case we consider the coupling of the GB
corresponding to the pion, $K$, $\eta$ and $\eta'$ mesons being inversely
proportional to the square of their respective masses, we find that
their couplings are of the order $a \alpha^2 \sim 0.02$,
$a \beta^2 \sim 0.02$ and $a \zeta^2 \sim 0.001$ for $a \sim 0.13$
which strangely agrees
with our values obtained through the fit. These findings are also in
agreement with the suggestions of Cheng and Li \cite{cheng} who have
advocated that the $\eta'$ contribution corresponds to the non-planar
contributions  in the $1/N_c$ expansion.

To summarize,  we have
investigated in detail the implications of the latest data pertaining to 
$\bar u-\bar d$ asymmetry and the spin polarization functions on the 
singlet Goldstone Boson
$\eta'$ within $\chi$CQM with configuration mixing for explaining the
``proton spin problem''. 
We find that the lower values of $\zeta$ are preferred
over the higher values.  Specifically, in the case of 
$\Delta_3$, $\Delta_8$, $\bar u-\bar d$,  $\bar u/\bar d$ and
$f_3/f_8$, it seems that the small non-zero  value of
$\zeta$ is preferred over 
$\zeta=0$.

\vskip .2cm
 {\bf ACKNOWLEDGMENTS}\\
The authors would like to thank S.D. Sharma
for a few useful discussions.
H.D. would like to thank CSIR, Govt. of India, for
 financial support and the chairman,
 Department of Physics, for providing facilities to work
 in the department. JMSR would like to thank
ICTP, Trieste, Italy, for hospitality
and financial support.

\pagebreak 

\begin{table}
\begin{center}
\begin{tabular}{|cccccc|}      \hline
Parameter$\rightarrow$ &$\phi$ &$a$ &$\alpha$ &$\beta$ &$\zeta$ \\  
\hline
Value & $20^{o}$ & 0.1 & 0.4 & 0.7 & $-0.3-\beta/2$  \\ 
& $20^{o}$ & 0.13 & 0.4 & 0.4 & $0.15-\beta/2$  \\ 
& $20^{o}$ & $0.354/(3-\beta)$ & 0.4 & 0.4 & 0  \\ \hline

\end{tabular}
\end{center}
\caption{ Input parameters and their values used in the analysis.} 
\label{input}
\end{table}

\begin{table}
\begin{center}
\begin{tabular}{|ccccc|}       \hline
Parameter & Data  & \multicolumn{3}{c|}{$\chi$CQM$_{{\rm config}}$}
\\  \cline{3-5} 
&      & $a=0.1$ & $a=0.14$ & $a=0.13$\\ 
&      & $\zeta=-0.65$ & $\zeta=0$ & $\zeta=-0.10$\\ \hline

$\Delta u$ & 0.85 $\pm$ 0.05 \cite{smc} & 0.95 & 0.91 & 0.91 \\
$\Delta d$ & $-$0.41  $\pm$ 0.05 \cite{smc}   & $-$0.31 
& $-$0.35& $-$0.36 \\
$\Delta s$ &$-$0.07  $\pm$ 0.05 \cite{smc} &$-$0.02&$-$0.02 &$-$0.02 \\ 
$\Delta_3$ & 1.267 $\pm$ 0.0035 \cite{PDG} & 1.27 & 1.26 & 1.27 \\ 
$\Delta_8$ & 0.58  $\pm$ .025 {\cite{PDG}} &0.67 & 0.60 & 0.59 \\ 
$\Delta \Sigma$ & 0.19  $\pm$ .025 {\cite{PDG}}& 0.31 &0.27 & 0.28 \\
\hline
\end{tabular}
\end{center}
\caption{The phenomenological values of the spin polarizations and 
dependent parameters.} \label{spin}
\end{table}

\begin{table}
\begin{center}
\begin{tabular}{|ccccc|}       \hline 
               
Parameter & Data  & \multicolumn{3}{c|}{$\chi$CQM} \\ \cline{3-5}  
&      & $a=0.1$ & $a=0.14$& $a=0.13$ \\  
&      & $\zeta=-0.65$ & $\zeta=0$ & $\zeta=-0.10$ \\ \hline 
$\bar u$ &-  &  0.168 &0.25 & 0.23 \\
 
$\bar d$ & - &  0.288 & 0.366& 0.35 \\
 
$\bar s$ &-   &   0.108 & 0.07& $0.07$ \\
 
$\bar u-\bar d$ & $-0.118 \pm$ .015 \cite{e866} & $-0.108$& $-0.116$ & $-0.117$ \\ 

$\bar u/\bar d$ & 0.67 $\pm$ 0.06 {\cite{e866}}  & 0.58 &0.68& 0.67  \\

$I_G$ & 0.254  $\pm$ .005  & 0.253 &0.255 & 0.255  \\
    
$f_u$ &-   &   0.655 &0.677& 0.675 \\

$f_d$ &- &  0.442 &0.470 & 0.466 \\

$f_s$ &  0.10 $\pm$ 0.06 {\cite{ao}}  &  0.061 &0.039 & 0.039  \\

$f_3$  &- & 0.213 & 0.207 & 0.209\\

$f_8$  &- & 0.975 & 1.07& 1.06 \\

$f_3/f_8$ & 0.21 $\pm$ 0.05 {\cite{cheng}}  &  0.22 & 0.19 & 0.20 \\ \hline

\end{tabular}
\end{center}
\caption{The quark flavor distribution 
functions and dependent parameters.}  
\label{quark}
\end{table}


\begin{thebibliography}{99}

\bibitem{manohar}  A. Manohar and H. Georgi, Nucl. Phys.
{\bf B 234}, 189 (1984).

\bibitem{wein} S. Weinberg, Physica {\bf A 96}, 327 (1979).

\bibitem{eichten}  E.J. Eichten, I. Hinchliffe and C. Quigg,
 Phys. Rev. {\bf D 45}, 2269 (1992).

\bibitem{cheng}  T.P. Cheng and Ling Fong Li, Phys. Rev. Lett.
{\bf 74}, 2872 (1995); hep-ph/9709293; Phys. Rev. {\bf D 57}, 344 (1998).

\bibitem{song} X. Song, J.S. McCarthy and H.J. Weber, Phys. Rev.
{\bf D 55}, 2624 (1997); X. Song, Phys. Rev. {\bf D 57}, 4114 (1998).

\bibitem{johan}  J. Linde, T. Ohlsson and Hakan Snellman, Phys. Rev.
{\bf D 57}, 452 (1998);
T. Ohlsson and H. Snellman, Eur. Phys. J., {\bf C 7}, 
501 (1999).


\bibitem{emc} EMC Collaboration, J. Ashman {\it et al.}, 
Phys. Lett. {\bf  206B}, 364 (1988); Nucl. Phys. {\bf B 328}, 1 (1989).

\bibitem{smc} SMC Collaboration, B. Adeva {\it et al.},  
Phys. Lett. {\bf  302B}, 533 (1993);
P. smc {\it et al.}, Phys. Rev. D {\bf 56}, 5330 (1997).

\bibitem{abe} E142 Collaboration, P.L. Anthony {\it et al.}, 
 Phys. Rev. Lett. {\bf 71}, 959 (1993);
 E143 Collaboration, K. Abe {\it et al.}, 
 Phys. Rev. Lett. {\bf 75}, 391 (1995).

\bibitem{hermes} 
HERMES Collaboration, K. Ackerstaff  {\it et al.}, 
Phys. Lett. {\bf 404B}, 383 (1997). 

\bibitem{nmc}  New Muon Collaboration, P. Amaudruz {\it et al.},
 Phys. Rev. Lett. {\bf 66}, 2712 (1991); M. Arneodo {\it et al.},
 Phys. Rev. {\bf D 50}, R1 (1994).

\bibitem{e866} E866/NuSea Collaboration, E.A. Hawker {\it et al.},
 Phys. Rev. Lett. {\bf 80}, 3715 (1998); J.C. Peng {\it et al.},
 Phys. Rev. {\bf D 58}, 092004 (1998); R. S. Towell {\it et al.},
{\it ibid.} {\bf 64}, 052002 (2001).

\bibitem{GSR}  K. Gottfried, Phys. Rev. Lett. {\bf 18}, 1174 (1967).

\bibitem{dgg} A. De Rujula, H. Georgi and S.L. Glashow,
Phys. Rev. {\bf D 12}, 147 (1975).

\bibitem{Isgur} N. Isgur, G. Karl and R. Koniuk, Phys. Rev. Lett.
{\bf 41}, 1269 (1978);
N. Isgur and G. Karl,  Phys. Rev. {\bf D 21}, 3175 (1980);
N. Isgur {\it et al.},  Phys. Rev.  {\bf D 35}, 
1665 (1987); P. Geiger and N. Isgur,  Phys. Rev. {\bf D 55}, 
299 (1997).

\bibitem{yaouanc}  A. Le Yaouanc, L. Oliver, O. Pene and J.C. Raynal,
 Phys. Rev. {\bf D 12}, 2137 (1975);  {\it ibid.}. {\bf 15}, 844 (1977).

\bibitem{riska} L.Ya. Glozman and D.O. Riska, Phys. Rep. {\bf 268},
263 (1996), L.Ya. Glozman, Z. Papp and W. Plessas, Phys. Lett. {\bf
381B}, 311 (1996).
 
\bibitem{chengspin} T.P. Cheng and Ling Fong Li, hep-ph/9811279.

\bibitem{prl} Adam P. Szczepaniak and Erie S. Swanson  Phys. Rev. Lett.    
{\bf 87}, 072001 (2001).


\bibitem{hd} H. Dahiya and M. Gupta, Phys. Rev. {\bf D 64}, 014013 (2001);
{\it ibid.} {\bf D 66}, 051501(R) (2002); 
{\it ibid.} {\bf D 67}, 074001 (2003);
{\it ibid.} {\bf 67}, 114015 (2003).

\bibitem{cheng1}  T.P. Cheng and Ling Fong Li, Phys. Rev. {\bf D 80},
 2789 (1998).

\bibitem{cg} J. Franklin, Phys. Rev. {\bf  182}, 1607 (1969).

\bibitem{psp} ``Proton spin crisis'' is not only limited to the spin
polarization functions but also to the flavor structure of
the nucleon, therefore ``proton spin crisis'' will be
referred to as ``PROTON SPIN PROBLEM''.

\bibitem{bass} Steven D. Bass, Phys. Lett. {\bf 463B}, 286 (1999);
{\it ibid.} Nucl. Phys. Proc. Suppl.  {\bf 105}, 56 (2002).

\bibitem{full} P.N. Pandit, M.P. Khanna and M. Gupta, 
J. Phys. G {\bf 11}, 683 (1985).

\bibitem{mgupta1} M. Gupta and N. Kaur,
Phys. Rev. {\bf D 28}, 534 (1983);  M. Gupta, J. Phys. G:
Nucl. Phys. {\bf 16}, L213 (1990).

\bibitem{yaoubook}  A. Le Yaouanc {\it et al.}, 
{\it Hadron Transitions in the Quark Model}, Gordon and Breach, 1988.

\bibitem{bjorken} J.D. Bjorken, Phys. Rev. {\bf 148}, 1467 (1966);
Phys. Rev.  {\bf D 1}, 1376 (1970).

\bibitem{ellis} J. Ellis and R.L. Jaffe, Phys. Rev.  {\bf D 9}, 1444 (1974);
{\it ibid.} {\bf 10}, 1669 (1974).

\bibitem{baldit}  NA51 Collaboration, A. Baldit {\it et al.},  Phys. Lett.
{\bf 253B}, 252 (1994).

\bibitem{neu charge} 
 M. Gupta and A.N. Mitra, Phys. Rev. {\bf D 18}, 1585 (1978); 
N. Isgur, G. Karl and D.W.L. Sprung,  {\it ibid} {\bf 23}, 163 (1981).

\bibitem{hdgluon} H. Dahiya and M. Gupta, hep-ph/0305327 
(to appear in Int. Jol. of Mod. Phys. A).

\bibitem{PDG}  K. Hagiwara {\it et al.}, 
Phys. Rev.  {\bf D 66}, 010001 (2002).

\bibitem{ao} A.O. Bazarko {\it et al.}, Z. Phys {\bf C 65}, 189 (1995); 
J. Grasser, H. Leutwyler and M.E. Saino, Phys. Lett. {\bf  253B}, 252 (1991);
S.J. Dong  {\it et al.}, Phys. Rev. Lett. {\bf 75}, 2096 (1995).

\end{thebibliography}
\end{document}